\documentclass{article} 
\usepackage{iclr2026_conference,times}

\usepackage{amsmath,amsfonts,bm}

\def\eqref#1{equation~\ref{#1}}

\def\1{\bm{1}}

\DeclareMathAlphabet{\mathsfit}{\encodingdefault}{\sfdefault}{m}{sl}
\SetMathAlphabet{\mathsfit}{bold}{\encodingdefault}{\sfdefault}{bx}{n}

\usepackage{hyperref}
\usepackage{xurl} 
\usepackage{graphicx}

\title{Efficiency vs Demand in AI Electricity: 

Implications for Post-AGI Scaling}

\author{Doyi Kim, Jiseok Ahn, Haewon McJeon, Changick Kim\thanks{Corresponding author.}\\
Korea Advanced Institute of Science and Technology (KAIST)\\
\texttt{\{doyi.kim, jiseok.ahn, hmcjeon, changick\}@kaist.ac.kr}
}

\iclrfinalcopy 
\begin{document}

\maketitle

\begin{abstract}
As AI capabilities and deployment accelerate toward a post-AGI era, concerns are growing about electricity demand and carbon emissions from AI computing, yet it is rarely represented explicitly in long-term energy–economy–climate scenario models. In such a setting, digital infrastructure scaling may be constrained by power-system dynamics. We introduce an AI computing sector into the Global Change Analysis Model (GCAM) and run U.S. scenarios that couple AI service growth with time-varying compute energy intensity and economic drivers. We find that service growth does not translate linearly into electricity demand: outcomes depend on efficiency trajectories and demand responsiveness. With sustained efficiency improvements, AI electricity demand remains moderated; with slower or saturating gains, income-driven demand dominates by mid-century. Sensitivity analyses show weak responsiveness to price signals but strong dependence on income growth, implying limited leverage from price-based mechanisms alone. Rather than offering a single forecast, we map conditions under which efficiency-dominant versus demand-dominant regimes emerge, providing a compact template for long-run AI electricity-demand scenarios and their implications for power-sector emissions.
\end{abstract}

\section{Introduction}
Artificial intelligence (AI) has rapidly transitioned from theoretics to a general-purpose technology embedded across economic sectors, driving rapid growth in large-scale data centers for training and inference. A recent report (\cite{iea2025energyai}) shows that data centers accounted for approximately 1.5\% of global electricity consumption in 2024, with usage growing at a 12\% annual rate over the past five years, raising concerns about grid reliability, energy affordability, and carbon emissions.

At the same time, the AI stack has experienced rapid improvements in efficiency. Historical trends in GPU performance for leading AI accelerators suggest that computational efficiency has improved by roughly 1.3$\times$ per year (\cite{pilz2025trends}), creating a countervailing dynamic in which growing AI service demand may be partially offset by declining energy per unit of compute. This tension has generated substantial uncertainty in projections of AI-driven electricity demand, with some studies warning of large increases in emissions (\cite{aljbour2024powering, bcg2025powerofcompute, gs2024generationalgrowth, shehabi2024usdcenergy}) and others arguing demand may remain manageable in the near term (\cite{nadel2025dcEfficiencyFlex, koomey2025guideperplexed, masanet2020recalibrating}). Continued improvements in specialized accelerators and data-center operations further complicate long-run projections.

Building on these near-term perspectives, we extend the analysis to mid-century and identify the conditions, especially efficiency persistence and demand elasticities, under which projections shift between efficiency-dominant and demand-dominant regimes. A macro-scale energy–economy model makes this interaction explicit by propagating these assumptions through power-sector transitions and emissions. In a potential post-AGI setting where AI services become pervasive across the economy, these long-run sensitivities matter because power-system dynamics may become a first-order constraint on digital infrastructure scaling. Accordingly, we embed AI computing in GCAM and run mid-century scenarios to map regime thresholds for AI electricity demand.

\begin{figure*}[t!]
 \centering
\includegraphics[width=1.0\linewidth]{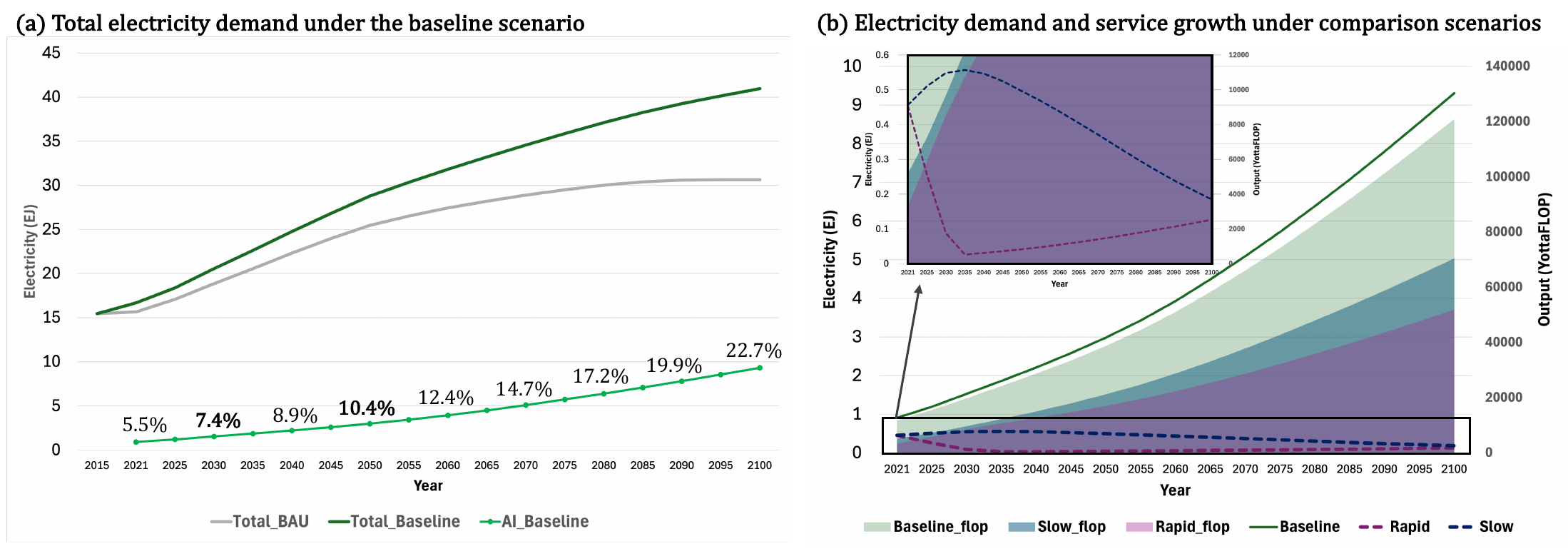} \caption{\textbf{AI-driven electricity demand under alternative efficiency assumptions.} (a) Total U.S. electricity demand under the business-as-usual(BAU) and baseline scenario that adds an AI data-center load with fixed compute energy intensity $\gamma$.
(b) AI data-center electricity demand (left axis, solid lines) and AI service output (right axis, shaded area) under alternative efficiency trajectories (\textit{Rapid} and \textit{Slow}). Service output grows rapidly in all scenarios, but electricity demand diverges depending on the assumed pace of efficiency improvement.}
 \label{fig:fig1}
\end{figure*}

\section{Experimental Set up}
We extend the Global Change Analysis Model (GCAM, \cite{calvin2019gcam}) by adding an explicit AI computing sector within the industrial module. GCAM is a partial-equilibrium, market-based macro-scale energy-economy model that links socioeconomic drivers, energy supply and demand, land use, and emissions. However, it does not explicitly represent AI-related computing demand, motivating a dedicated sectoral representation.

We measure AI service output as floating-point operations (FLOP), a proxy for computational workload (\cite{pilz2025trends}). AI electricity demand is modeled as the product of service output $S(t)$ (FLOP) and a system-level compute energy intensity coefficient $\gamma(t)$ (Joule/FLOP)\footnote{We express $S(t)$ in YottaFLOP (YF, $1~\mathrm{YF}=10^{24}$ FLOP), so $\gamma(t)$ is reported in EJ/YF (equivalently J/FLOP).}, such that
\[
E(t) = S(t)\times\gamma(t),
\]
where $\gamma(t)$ captures chip-level efficiency as well as data center overheads (cooling, networking, storage, and power usage effectiveness). In GCAM, $S(t)$ is determined endogenously from demand elasticities and macroeconomic drivers. We consider a baseline scenario with a fixed $\gamma$ ($7.7\times10^{-5}$ EJ/YF) and two alternative efficiency trajectories: \textit{Rapid} (doubling compute performance roughly every 2.34 years until 2035) and \textit{Slow} (roughly 180\% per decade). We also run sensitivity scenarios that vary price elasticity (-0.2, -0.7, -1.2) and income elasticity (1.6, 2.5, 3.5) around the baseline (Table~\ref{tab:table1}). Full parameter and implementation details are provided in Appendix~\ref{app:param}.

\begin{table*}[htbp]
  \centering
  \caption{Scenarios used in this study}
  \label{tab:table1}
  \includegraphics[width=1.0\linewidth]{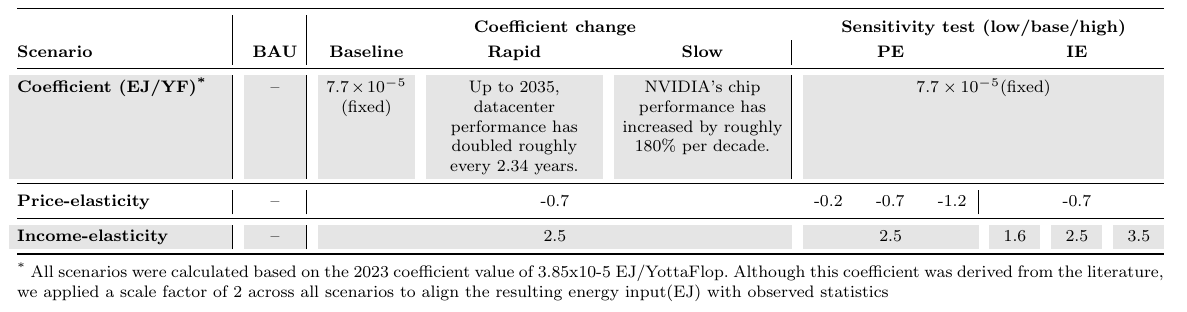}
\end{table*}

\section{Results}
\subsection{Baseline Impacts in GCAM}
Under the U.S. baseline scenario, annual electricity demand from AI data centers reaches approximately 1.5 EJ (about 420 TWh) by 2030 and 3 EJ (about 830 TWh) by 2050, accounting for roughly 10\% of total U.S. electricity use in 2050 (broadly consistent with recent U.S. estimates (\cite{bcg2025powerofcompute, aljbour2024powering})). These near-term magnitudes are also comparable to recent short-horizon estimates (e.g., \cite{kumar2025trends}), providing an external anchor for our baseline calibration.

Incorporating time-varying efficiency, as in the \textit{Rapid} and \textit{Slow} efficiency scenarios, substantially moderates electricity demand relative to the baseline (Fig.~\ref{fig:fig1}(b)). The \textit{Rapid} case drives demand to implausibly low levels and is therefore treated as an upper-bound stress test for regime boundaries, while the \textit{Slow} case still reduces demand but by a more limited margin. Put simply, electricity demand reflects how much compute is used and how much energy each unit of compute requires; efficiency trajectories determine whether the latter can offset the former.

\subsection{Demand Responsiveness of AI Computing}
\begin{figure*}[t!]
 \centering
\includegraphics[width=0.9\linewidth]{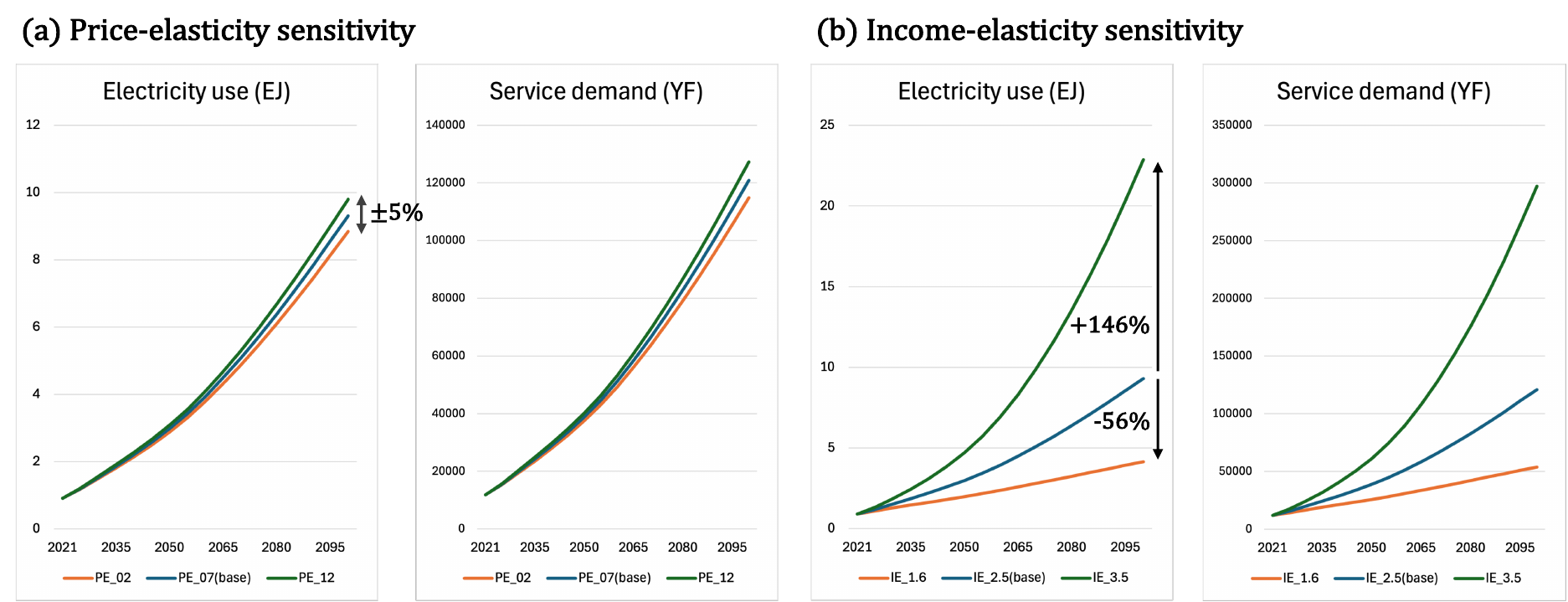} \caption{\textbf{Sensitivity of AI computing electricity demand to price and income elasticitie.} (a) Electricity demand and AI service output under alternative price elasticity assumptions (All other parameters are the same as baseline). (b) Electricity demand and AI service output under alternative income elasticity assumptions.}
 \label{fig:fig2}
 \vspace{-3mm}
\end{figure*}

To quantify this, we conduct a set of sensitivity experiments to isolate the role of demand responsiveness in determining how rapidly AI service growth translates into electricity use. In Fig.~\ref{fig:fig2}(a), varying the price elasticity has a relatively modest effect on both electricity use and service output. Across the tested range, changes in electricity demand remain within ±5\% relative to the baseline. This price inelasticity is consistent with the operational properties of AI stacks, including high fixed capital costs and continuous operation.

In comparison, Fig. \ref{fig:fig2}(b) shows that demand is highly sensitive to assumptions about income elasticity. Increasing the income elasticity of AI computing demand leads to large and nonlinear increases in both demand and output. 
Under higher income elasticity assumptions (IE\_3.5 scenario), mid-century electricity demand can increase about 150\% relative to the baseline elasticity case, suggesting income-driven expansion as a dominant driver of long-term electricity use. As income rises, AI adoption expands not only in intensity but also in scope, penetrating new applications and sectors. This behavior differs markedly from that of traditional consumer services and underscores the importance of modeling AI demand as an intermediate input within energy-economy scenario models.

\subsection{Efficiency Trajectories and Regime Thresholds}
Next, we examine how this demand behavior interacts with alternative efficiency trajectories. We ask when demand outweighs efficiency improvements in long-term electricity use. In a post-AGI setting where AI services diffuse broadly across sectors, these regime boundaries help identify when power-system constraints are likely to become a first-order limiter of compute scaling.

\begin{figure*}[t!]
 \centering
\includegraphics[width=0.9\linewidth]{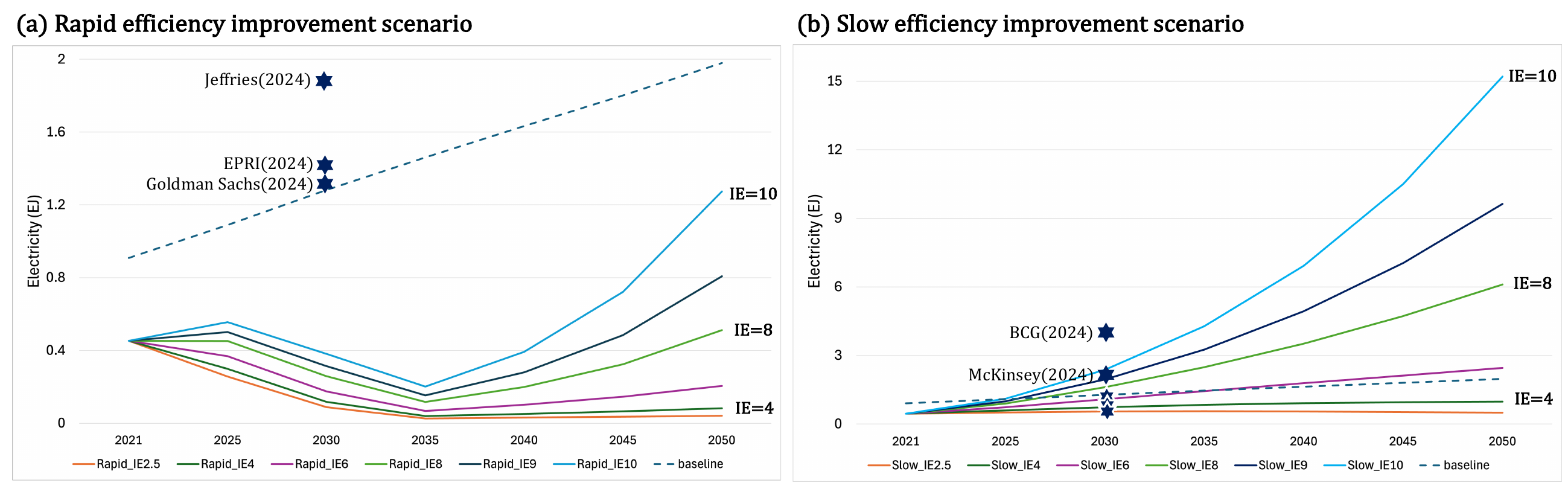} \caption{\textbf{Income elasticity thresholds under alternative efficiency trajectories.} Electricity demand responses to varying income elasticity assumptions under (a) \textit{Rapid} and (b) \textit{Slow} efficiency improvement scenarios. In (a), demand growth dominates only under very high income elasticities, whereas in (b) it does so at more moderate values. Blue stars indicate 2030 U.S. electricity demand estimates from the \cite{iea2025energyai} (Table 3.2), highlighting the consistency with external projections.}
 \label{fig:fig3}
 \vspace{-3mm}
\end{figure*}

In Fig.~\ref{fig:fig3}(a), the \textit{Rapid} trajectory delays demand-driven increases in electricity use until efficiency gains plateau, whereas the \textit{Slow} trajectory allows income-driven growth to dominate earlier despite continued improvements. As a result, long-term projections depend not only on the pace of efficiency gains, but also on their persistence over time. Differences in efficiency trajectories, rapid early improvements followed by saturation versus slower, sustained improvements can shift the demand-dominant threshold by decades. This highlights the need to jointly consider demand response and the timing/durability of efficiency gains when representing AI compute in macro-scale energy and emissions scenario models.

\section{Discussion and Limitations}
The results suggest that rising AI demand does not translate mechanically into proportional increases in electricity use. Instead, long-term outcomes depend on the interaction between service growth and system-level efficiency. Under scenarios with sustained efficiency improvements, electricity demand growth can be substantially moderated, whereas slower or saturating efficiency gains allow income-driven demand to dominate over longer time horizons. In a potential post-AGI setting where AI services diffuse broadly across sectors, these regime boundaries matter because power-system dynamics may become a first-order constraint on the pace and carbon intensity of digital infrastructure scaling.

These findings may appear different from some recent reports that project rapid growth in data center electricity demand. We do not interpret this divergence as a contradiction, but as reflecting differences in scope and assumptions—particularly how quickly compute energy intensity improves and whether efficiency gains persist or plateau. By embedding AI-driven electricity demand within a macro-scale energy–economy framework and extending scenarios to mid-century, our analysis makes these interactions explicit and identifies conditions under which projections can diverge.

Sensitivity experiments for the U.S. indicate weak responsiveness to price signals but strong dependence on income growth, implying limited leverage from price-based mechanisms alone and a first-order role for efficiency progress and the scale of service expansion. Because these elasticities are calibrated and evaluated for the U.S. setting, the relative importance of price versus income effects should be validated across other countries with different market structures, electricity pricing regimes, and digital adoption pathways.

Several limitations apply. AI service output is determined endogenously within GCAM from assumed demand elasticities and macroeconomic drivers, rather than imposed as an arbitrary time series. The AI computing sector introduced here is a simplified representation and does not capture heterogeneity in workloads, infrastructure design, or deployment strategies. Moreover, we focus on the joint dynamics of service growth and system-level efficiency; location-specific grid constraints and generation-mix heterogeneity that shape siting decisions and emissions are left for future work.


\subsubsection*{Acknowledgments}
We thank Professor Sonny Kim and Artem Vasilev for helpful discussions and guidance on the scenario setup used in this study.

\bibliography{iclr2026_conference}
\bibliographystyle{iclr2026_conference}

\newpage
\appendix
\section{Appendix}\label{app:param}
\subsection{AI computing sector implementation}\label{subsec2}

\begin{figure*}[htbp]
 \centering
\includegraphics[width=0.7\linewidth]{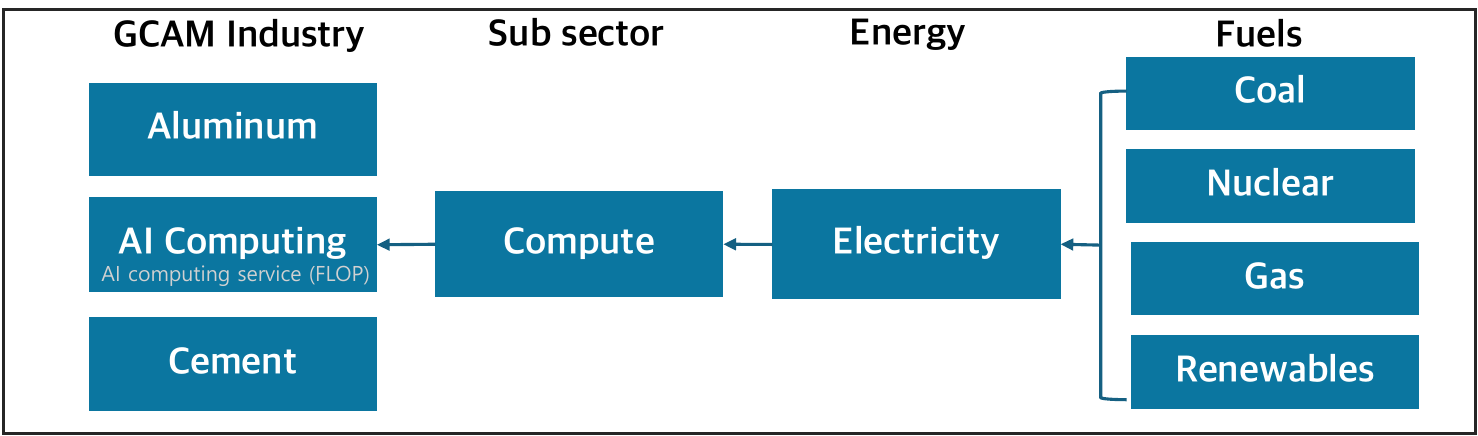} \caption{Schematic representation of the AI computing sector integrated into GCAM}
 \label{fig:fig6}
 \vspace{-3mm}
\end{figure*}

We extend the Global Change Analysis Model (GCAM) by introducing an explicit AI computing sector within the industrial module. In its standard configuration, GCAM represents industrial energy demand through aggregated sectors such as cement, aluminum, and chemicals. These representations are insufficient to capture the distinctive characteristics of AI-related electricity demand, including computation-based service output, rapid efficiency change, and continuous operation. To address this gap, we introduce AI computing as a new industrial service sector that is structurally comparable to existing GCAM industries, but differs in its definition of output and its linkage to electricity demand.

The AI computing sector is implemented as a new subsector under the GCAM industry module (Fig.~\ref{fig:fig6}). This subsector produces an abstract service, AI computation, that is demanded by the economy and supplied using electricity as the energy input. By embedding the sector within GCAM’s market structure, AI electricity demand competes with other end-use sectors for electricity supply, allowing power-system responses and emissions to be determined endogenously.

Note that AI service output is not imposed as an arbitrary time series in GCAM. Instead, it is determined endogenously from the assumed demand elasticities and macroeconomic drivers through the model’s market equilibrium. We use the baseline (fixed $\gamma$) as a reference case anchored to near-term U.S. estimates, and treat alternative efficiency paths as counterfactual scenarios to map long-run sensitivities.

\paragraph{Defining AI service output}
We define AI service output in terms of floating-point operations (FLOP). This choice allows AI demand to be linked directly to hardware efficiency trends and electricity consumption (\cite{pilz2025trends, desislavov2023trends, tripp2024measuring}). Service output is expressed at large numerical scales using PetaFLOP ($10^{15}$) and YottaFLOP ($10^{24}$) units to ensure numerical stability within the GCAM solution algorithm.

\paragraph{Linking computation to electricity demand}
Electricity demand from AI computing is derived by converting service output(\textit{S}(t)) into energy use through a system-level efficiency coefficient. Total electricity demand E(t) is calculated as:
\[E(t) = S(t)\times \gamma(t)\]

where $\gamma(t)$ represents the electricity required per unit of computation (J/FLOP). This coefficient captures not only chip-level efficiency, but also system-level factors such as cooling, networking, storage, and power usage effectiveness.

\subsection{Parameterization of AI computing demand}
To represent AI computing within GCAM, we parameterize demand using a small set of key assumptions governing service growth, price responsiveness, energy efficiency, and non-energy input costs. Rather than modifying GCAM’s underlying functional forms, we retain the standard industrial sector formulation and focus on constraining these parameters using ranges informed by the existing literature. Exact parameter values used in each scenario are reported in the Table \ref{tab:table1}.

\paragraph{Income elasticity}
Income elasticity governs how AI computing service demand responds to economic growth. Given that AI services are primarily used for information processing, digital services, education, and entertainment, we adopt income elasticity values in the range of \textit{1.6–2.0} based on estimates for comparable service-oriented sectors (\cite{bekkers2025lookingglass, liu2024generativeai}). To explore more aggressive AI adoption trajectories characterized by rapid service expansion, we additionally consider higher values up to \textit{2.5}.

\paragraph{Price elasticity}
Price elasticity captures the sensitivity of AI computing demand to changes in service price. Consistent with prior studies of internet-based and digitally mediated services, we assume relatively low price responsiveness. Price elasticity values are set between $-0.2$ and $-0.7$, reflecting the small share of electricity and compute costs in the overall value of AI services and the limited role of price signals in constraining demand (\cite{goel2006demand, shiva2025bitcoin}).

\paragraph{Efficiency coefficient}
Energy efficiency links AI service output to electricity use and is expressed as energy consumption per unit of computation (EJ/YottaFLOP). Baseline efficiency values ($3.83 \times 10^{-5}\ \mathrm{EJ/YottaFLOP}$) are derived from recent estimates of GPU and data center performance in 2023 (\cite{benhari2024green, shehabi2024usdcenergy, desislavov2023trends}), and alternative trajectories are constructed to represent constant efficiency, \textit{rapid} and \textit{slow} improvement scenarios. In the \textit{rapid} improvement case, efficiency gains accelerate in the near term and saturate after 2035 (\cite{hobbhahn2022predictinggpu}), whereas in the \textit{slow} improvement case, efficiency improves steadily through the end of the century.

\paragraph{Non-energy cost}
Non-energy costs include capital investment, labor, and operational expenditures associated with AI data centers. Using published estimates of U.S. data center capital expenditures (\cite{gs2024generationalgrowth}), we derive an implied non-energy cost of approximately 0.0015 1975\$/PetaFLOP of computational output. As a robustness check, we also consider alternative assumptions in which non-energy costs are set to two to four times energy costs, yielding comparable magnitudes.
\newline

Together, these parameters define the demand and cost structure of the AI computing sector while remaining consistent with the broader GCAM framework.

\end{document}